# Distributed Dynamic State-Input Estimation for Power Networks of Microgrids and Active Distribution Systems with Unknown Inputs

Bang L. H. Nguyen, Tuyen V. Vu, Joseph M. Guerrero, Mischael Steurer, Karl Schoder, and Tuan Ngo

*Abstract*—This paper proposes a joint input and state dynamic estimation scheme for power networks in microgrids and active distribution systems with unknown inputs. The conventional dynamic state estimation of power networks in the transmission system relies on the forecasting methods to obtain the state-transition model of state variables. However, under highly dynamic conditions in the operation of microgrids and active distribution networks, this approach may become ineffective as the forecasting accuracy is not guaranteed. To overcome such drawbacks, this paper employs the power networks model derived from the physical equations of branch currents. Specifically, the power network model is a linear state-space model, in which the state vector consists of branch currents, and the input vector consists of bus voltages. To estimate both state and input variables, we propose linear Kalman-based dynamic filtering algorithms in batch-mode regression form, considering the cross-correlation between states and inputs. For the scalability of the proposed scheme, the distributed implementation is also presented. Complementarily, the predicted state and input vectors are leveraged for bad data detection. Results carried out on a 13-bus microgrid system in real-time Opal-RT platform demonstrate the effectiveness of the proposed method in comparison with the traditional weighted least square and tracking state estimation methods.

*Index Terms*-- Bad data detection, dynamic state estimation, distributed scheme, active distribution networks, Kalman filter, input estimation, micro-PMU, microgrids.

B. L. H. Nguyen and T. V, Vu are with Clarkson University, Potsdam, NY, USA (e-mail: nguyenbl@clarkson.edu, tvu@clarkson.edu ).
J. M. Guerrero is with the Center of Research on Microgrids (CROM), Department of Energy Technology, Aalborg University,9220 Aalborg East, Denmark (Tel: +45 2037 8262; Fax: +45 9815 1411; e-mail: joz@et.aau.dk ).
M. Steurer and K. Schoder are with the Center of Advanced Power Systems, Florida State University, Tallahassee, FL 32310 USA (e-mail: steurer@caps.fsu.edu, schoder@caps.fsu.edu ).
Tuan Ngo is with Electric Power Engineers, Austin, TX, USA (email: ngotuan@utexas.edu)

I. INTRODUCTION

STATE estimation (SE) and bad data detection (BDD) play a crucial role in the power and energy management of power systems. They have a tight supporting relationship with other tools such as topology processing, load forecasting, economic dispatch, load-frequency control, contingency analysis, and security assessment [1].

Conventionally, SE and BDD algorithms are accomplished in a static and centralized manner with the assumption of quasi-steady conditions which allow techniques such as weighted least squares (WLS), least absolute value (LAV), least median of squares, least trimmed squared, and generalized maximum-likelihood (GML) to be applied. Static SE (SSE) only concern the measurement models to derive their formulations. These algorithms process the measurements in a single scan and restart with new data regardless of the previous estimates. Besides, they require redundant data to filter out measurement errors effectively and are unable to detect false data injection attacks (FDIA) [2].

To address these problems, recent studies focus on dynamic state estimation (DSE) using Kalman-based filtering methods [1]. Unlike SSE, DSE requires a dynamic model where a state-transition model is specified, then DSE can obtain the current estimates using previous estimates and current measurements [3]. The conventional state-transition equations rely on a forecasting model of the bus voltages [4]. Most simply, the future bus voltages are forecasted as the same as those of the previous time step in tracking state estimation (TSE) [5]. This approach is effective in a stiff grid in power transmission systems, where bus voltages are less sensitive to load changes. However, in weak grids such as microgrids and active distribution systems, voltage variations are sensitive to load changes. Importantly, with the rising integration of renewable-based distributed generation units (DGU) and prosumer loads, the behavior of bus voltages can be highly dynamic with large variance [6]. Under these conditions, forecasting accuracy is not guaranteed.

As a result, many published SE algorithms for the distribution system still focus on static methodologies [7]. In detail, they put research efforts on the formulation of static distribution systems such as nodal-voltage or branch-current based [8], three-phase based SEs [9], *pseudo* measurement generations [10], and also decentralized implementation [11], [12]. For DSE, several regression-based forecasting methods are adapted under certain assumptions of operating conditions. [13] implements the classical TSE model along with a sequential discrete Kalman filter (KF) in a field-programmable-gate-array (FPGA) prototype for active distribution networks. In [14], the first-order AutoRegressive model AR(1) with the coefficient of one is used to build a load evolution model, then SE process is performed by an ensemble KF. Both [13] and [14] have to presume that the considered system is *quasi-static*. Dispensing this presumption, [15] employs Holt's short-term forecasting method [16] to predict the dynamic behavior of state vector and accomplishes the filtering process with extended and unscented KFs. In [17], the first-order prediction-correction method [18] is applied to formulate a time-varying optimization problem for distribution dynamic state estimation. The authors indicated that this methodology outperforms the traditional KF-based approaches in terms of flexibility and computational efficiency. It is worth pointing out that these mentioned DSE algorithms still rely on the forecasting methodologies. Alternatively, the physical model of the power grid can be utilized. However, the existed researches only consider specific cases [19], [20]. In addition, they assumed that the input vector is accurately available, whereas they may be measured with particular errors or may be unavailable.

To tackle the aforementioned issues, this paper proposes a joint input and state dynamic estimation scheme for power networks in microgrids and active distribution systems. The distinctive contributions of this paper are as follows:

1) The paper offers an unique scheme that yields the generality and flexibility in microgrid state and input estimation. The proposed scheme employs the physical equations of branch currents for the power network models. Combined with the measurement equations from micro-Phasor Measurement Unit (µPMU), a linear state-space representation is obtained with branch currents being the state vector and bus voltages being the input vector. Here, the dynamic model of power networks is fully decoupled from other components. If bus voltage models are unavailable or unreliable, they are considered as unknown inputs.

2) To the best of our knowledge, the paper proposes unknown-input Kalman-based dynamic filtering algorithms to simultaneously estimate both state and input vectors of power networks model in batch-mode regression form for the first time. The proposed algorithms allow for cross-correlations between states and inputs.

3) The proposed approach is scalable for large-scale power systems with a distributed implementation, where advanced assimilation procedures are derived and demonstrated. This distributed practice reduces the system complexity and the computational burden while assuring scalability.

4) Bringing to completion, a generalized effective bad data detection method based on Mahalanobis distance is compared and contrasted with existing techniques to detect a false data injection attack at the bus voltage measurements.

The rest of this paper is organized as follows. The state-space model formulations of the proposed scheme are presented in Section II. Section III presents the proposed Kalman-based filtering algorithms and bad data detection. In Section IV, the simulation results of case studies are provided and discussed. Finally, Section V concludes the paper.

## II. STATE-SPACE MODEL FORMULATIONS

In this Section, we first construct the general expression of the power networks model (A), which is decoupled from the rest of the system. The dynamic model of bus voltage (B) is also discussed. Then, the measurement equations (C) are considered. Combining dynamics models and measurement equations, a state-space representation (D) of power networks and available component models are formed. For distributed implementation, the state-space models are partitioned (E) followed by specific areas. Finally, remarks are discussed.

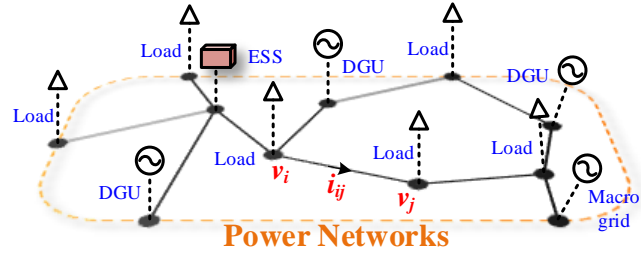

Fig. 1. Illustrated single-line diagram of microgrids and active distribution systems.

*A. General Expression of Power Networks Model*

A microgrid or an active distribution system includes power networks, DGU, loads, energy storage systems (ESS), and the connection to macro-grid, as shown in Fig. 1. We characterize the dynamic of power networks by the following differential equations of branch currents in *abc*-frame and *dq*-frame, respectively:

$$\frac{di_{ij,abc}}{dt} = -\frac{R_{ij}}{L_{ij}} i_{ij,abc} + \frac{1}{L_{ij}}(v_{i,abc} - v_{j,abc}), \quad (1)$$

$$\frac{di_{ij,dq}}{dt} = -\left(\frac{R_{ij}}{L_{ij}} + j\omega\right) i_{ij,dq} + \frac{1}{L_{ij}}(v_{i,dq} - v_{j,dq}), \quad (2)$$

where the subscripts *i*, *j* denote bus index; •$_{abc}$ denotes separately (•$_a$, •$_b$, •$_c$); •$_{dq}$ denotes (•$_d$ + j•$_q$); $i_{ij}$ is the branch current flow from bus *i* to bus *j*; $v_i$, $v_j$ is the bus voltage at bus *i*, *j*, respectively; $R_{ij}$ and $L_{ij}$ are the line resistance and inductance, respectively. The above *linear* differential equation of power networks is the *approximated* transient model of a medium-voltage distribution line, where the parameters are considered as fixed and lumped with negligible shunt capacitances [21]. This model has been widely utilized in many research papers [22], [23] addressing microgrid dynamic characterization.

A power network with *m* buses and *n* power lines can be considered as a directed graph having *m* vertices and *n* edges. Therefore, based on (2), the dynamic model of a power network is generally expressed as

$$\frac{d}{dt}\left[i_{ij,dq}\right]_{n\times 1} = -diag\left[\frac{R_{ij}}{L_{ij}} + j\omega\right]_{n\times n}\left[i_{ij,dq}\right]_{n\times 1} + \left[\frac{1}{L_{ij}}\right]_{n\times 1} \circ [B_{ind}^T]_{n\times m}[v_{i,dq}]_{m\times 1}, \quad (3)$$

where $\left[i_{ij,dq}\right]_{n\times 1}$ denotes the state vector of *n* branch currents; $[v_{i,dq}]_{m\times 1}$ denotes the unknown input vector of *m* bus voltages; (∘) indicates the element-wise product; the transposed incidence matrix $[B_{ind}^T]$ is defined as

$$[B_{ind}^T]_{n\times m} = \begin{bmatrix} b_1^{ij} & \cdots & b_m^{ij} \end{bmatrix}_{n\times m}, \quad (4)$$

with $b_k^{ij}, k \in [1, ..., m]$ indicates the direction of the branch current $i_{ij}$ to a bus *k* as

$$b_k^{ij} = \begin{cases} 1 \; if \; k = i \\ -1 \; if \; k = j \\ 0 \; if \; k \neq i,j \end{cases}. \tag{5}$$

*B. Bus Voltage Dynamic Models*

From the power networks model, the injected currents at a bus can be determined. Considering the dynamic models of a bus voltage, it should take the injected current $i_i$ as the input as:

$$v_{i,k+1} = f(v_{i,k}, i_{i,k}) + w_{v,k} \tag{6}$$

where $k$ is the time step, $w_{v,k}$ are the modeling uncertainties. This model depends on what components are connected to that bus, such as generation units, loads, ESS, or the macro-grid. Accurately modeling all these components are relatively impractical or requires certain presumptions. In this paper, if such dynamic models of bus voltages are unknown, then the corresponding bus voltages are considered as unknown inputs. If there is a bus voltage with a reliable dynamic model, it can be included in the power networks model.

*C. Measurement Considerations*

Measurement functions of PMU measurements for a bus voltage and a branch current can be given as follows:

$$\begin{bmatrix} z_{vi,d} \\ z_{vi,q} \end{bmatrix} = \begin{bmatrix} 1 & 0 \\ 0 & 1 \end{bmatrix} \begin{bmatrix} v_{i,d} \\ v_{i,q} \end{bmatrix} + e_{vi}, \tag{7}$$

$$\begin{bmatrix} z_{ij,d} \\ z_{ij,q} \end{bmatrix} = \begin{bmatrix} 1 & 0 \\ 0 & 1 \end{bmatrix} \begin{bmatrix} i_{ij,d} \\ i_{ij,q} \end{bmatrix} + e_{cij}, \tag{8}$$

where $z_{vi}$, $z_{ij}$ are the measurement values of branch currents and bus voltages with $e_{vi}$ and $e_{cij}$ being the measurement noises, respectively.

This paper focuses on capturing the dynamics of power networks according to the swift change of DGUs and loads at the timescale of around 10 *ms*. Therefore, only PMU measurements are considered here. Since, in practice, only PMU measurements may not be sufficient to ensure the observability of the state estimation process, smart meter data can also be incorporated into the linear formulation of state estimation as expressed [24]. The observability problems are considered in Section III-*D*.

*D. State-Space Representation*

Combining the available dynamic models and measurement equations, the discretized state-space representation can be assembled as

$$\begin{cases} x_k = Ax_{k-1} + Bu_{k-1} + w_{k-1} \\ z_{x,k} = Cx_k + v_{x,k} \\ z_{u,k} = Du_k + v_{u,k}, \end{cases} \quad (9)$$

where $k$ denotes the time step, $x_k \in \mathbb{R}^n$ is the state vector; the unknown input vector is $u_k \in \mathbb{R}^m$; $A$, $B$ are the state and input matrices; $C$, $D$ are measurement matrices; $w_{k-1}$ represents the modeling errors with covariance matrix $Q_{k-1}$; $z_{x,k} \in \mathbb{R}^p$ are the state measurements; $z_{u,k} \in \mathbb{R}^l$ are the input measurements; $v_{x,k}$ and $v_{u,k}$ are the measurement noises with covariance matrices $R_{x,k}$ and $R_{u,k}$, respectively.

*E. Model Partitioning*

For large power systems, the entire power networks can be partitioned into multiple areas with the single- or multi-terminal shared loads, as shown in Fig. 2. The local state-space model can be built as the same as the centralized model but only considering an area, as follows:

$$\begin{cases} x_{i,k} = A_i x_{i,k-1} + B_i u_{i,k-1} + w_{i,k-1} \\ z_{xi,k} = C_i x_{i,k} + v_{xi,k} \\ z_{ui,k} = D_i u_{i,k} + v_{ui,k}, \end{cases} \quad (10)$$

where $i$ indicates the area index. Since the branch-current equations only share the joint inputs of bus voltages, the local models are also decoupled. As illustrated in Fig. 2(a), when two areas share the single-terminal loads, their input vectors have the same variable of bus voltages at the points of coupling. Otherwise, their state and input vector has no variables in common, as shown in Fig. 2(b).

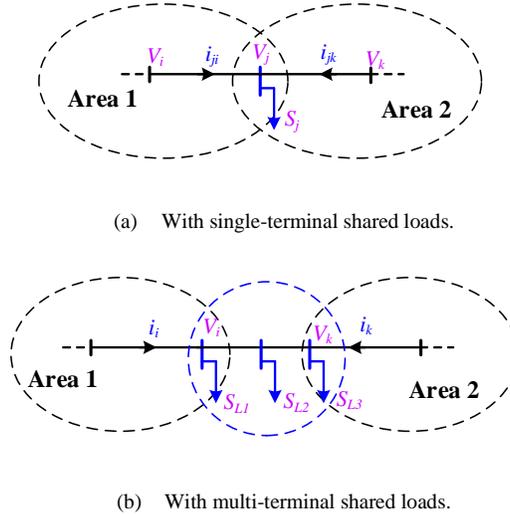

(a) With single-terminal shared loads.

(b) With multi-terminal shared loads.

Fig. 2. Multi-area partitioning of microgrids with the shared loads.

**Remark of decoupled dynamic models:** This approach was widely considered for decentralized DSE of synchronous generators in the transmission system in [25], [26]; since it alleviates the problem complexity and computational cost. Differently, this paper focuses on the decoupled dynamic models of power networks in microgrids and active distribution systems. Moreover, we propose joint states and unknown inputs dynamic estimation algorithms considering the cross-correlations between input and state vectors. The dynamic models of input variables, such as bus voltages in this paper, can be involved if reliably available, but it is not a must.

**Remark of µPMU measurements:** The distribution-level PMUs (µPMU), are more popular. Therefore, on one hand, it is reasonable and timely to design dynamic estimation techniques taking advantage of µPMU measurements [17]. On the other hand, the availability of effective µPMU-based estimation methods also appeals for deploying more such devices. In comparison, the PMU measurements are more accurate ($\pm 0.05\%$ total vector error - TVE). In addition, the updating interval of PMU-based measurements (about 0.4 *ms*) is much smaller than remote terminal unit (RTU)-based (about 2-4 *s*) and advanced metering infrastructure (AMI)-based *pseudo-* power flow measurements (about 15 minutes) [15].

### III. Joint State & Input Kalman-based Dynamic Filtering with Unknown Inputs

This Section presents the proposed Kalman-based dynamic filtering for joint state and input estimation. Here, we first analyze the centralized algorithm (A), which is deployed in a single estimator. Next, the distributed algorithm (B) is developed considering the communication between the neighboring estimators. Then, bad data detection (C) is proceeded by evaluating the integrity of predicted and actual measurements. Finally, observability (D) is considered.

#### A. Centralized Dynamic State-Input Estimation (DSIE)

From (9) or (10), the input-to-output relationship of the system can be derived as

$$z_{x,k} = CAx_{k-1} + CBu_{k-1} + Cw_{k-1} + v_{x,k}. \tag{11}$$

Combining (11) with the input measurement at $(k-1)$ yields a batch-mode regression form as

$$\begin{bmatrix} z_{x,k-1} \\ z_{u,k-1} \\ z_{x,k} \end{bmatrix} = \begin{bmatrix} C & 0 \\ 0 & D \\ CA & CB \end{bmatrix} \begin{bmatrix} x_{k-1} \\ u_{k-1} \end{bmatrix} + \begin{bmatrix} v_{x,k-1} \\ v_{u,k-1} \\ Cw_{k-1} + v_{x,k} \end{bmatrix}. \tag{12}$$

Solutions of (12) yield the estimates $\begin{bmatrix} \hat{x}_{k-1} \\ \hat{u}_{k-1} \end{bmatrix}$ and covariance matrix $P_{k-1} = \begin{bmatrix} P_{x,k-1} & P_{xu,k-1} \\ P_{ux,k-1} & P_{u,k-1} \end{bmatrix}$. Then, the predicted $\hat{x}_{k|k-1}$ and covariance can be achieved as

$$\hat{x}_{k|k-1} = A\hat{x}_{k-1} + B\hat{u}_{k-1} \tag{13}$$

$$P_{x,k|k-1} = [A \quad B]P_{k-1}\begin{bmatrix} A \\ B \end{bmatrix} + Q_{k-1}. \tag{14}$$

Hence, the updated estimates can be obtained as

$$\hat{x}_k = \hat{x}_{k|k-1} + K_k(z_{x,k} - C\hat{x}_{k|k-1}) \tag{15}$$

$$P_{x,k} = (I - K_k C)P_{k|k-1}, \tag{16}$$

where $K_k = P_{k|k-1}C^T(CP_{k|k-1}C^T + R_{x,k})^{-1}$ is the optimal Kalman gain.

The procedure of the centralized dynamic state-input estimation is summarized as follows.

---

**Algorithm 1:** Centralized Dynamic State-Input Estimation

---

**Data Requires**: $z_{x,k-1}, z_{u,k-1}, z_{x,k}$

1. Solve (12) for $\hat{x}_{k-1}$, $\hat{u}_{k-1}$ and $P_{k-1}$.

2. Predict $\hat{x}_{k|k-1}$ and $P_{x,k|k-1}$ using (13) and (14)

3. Update $\hat{x}_k$ and $P_{x,k}$ using (15) and (16)

4. Repeat step 1 at time step $k$

---

**Remarks**:

1) The batch-mode regression form of (12) can be solved using different methods such as WLS, LAV, or GML. This paper simply uses the WLS technique as a base-line solution. This yields the estimates

$$\begin{bmatrix}\hat{x}_{k-1}\\\hat{u}_{k-1}\end{bmatrix} = (\mathcal{O}^T R^{-1}\mathcal{O})^{-1}\mathcal{O}^T R^{-1}\begin{bmatrix}z_{x,k-1}\\z_{u,k-1}\\z_{x,k}\end{bmatrix}, \quad (17)$$

the weighted matrix $R_{k-1} = diag(R_{x,k-1}, R_{u,k-1}, E_{x,k-1})$, where $E_{x,k-1}$ is the covariance matrix of $Cw_{k-1} + v_{x,k}$, and $E_{x,k-1} = CQ_{k-1}C^T + R_{x,k}$. The estimated covariances are

$$P_{k-1} = \begin{bmatrix}P_{x,k-1} & P_{xu,k-1}\\P_{ux,k-1} & P_{u,k-1}\end{bmatrix} = (\mathcal{O}^T R^{-1}\mathcal{O})^{-1}, \quad (18)$$

with $\mathcal{O} = \begin{bmatrix}C & 0\\0 & D\\CA & CB\end{bmatrix}$.

2) The algorithm 1 is more complex than WLS and conventional methods since it includes steps similar to both WLS and standard Kalman filter. However, it is less complex than recent methods in [13], [14], [17], [18].

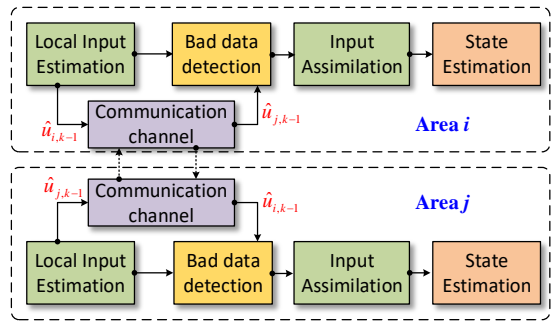

Fig. 3. Block diagram of local dynamic state-input estimation and communication with neighboring estimators.

## B. Distributed Implementation

Fig. 3 depicts the block diagram of the distributed implementation. There are five block functions in a distributed setting of a local estimator. First, the local input estimation is performed by solving (12) for the local area. Second, the voltage values at the shared buses is exchanged through a communication channel between the neighboring estimators. Third, bad data detection will check the consistency of local data and data received. The detailed analysis of this stage is presented in Section III-C. Fourthly, if there is no problem being detected, the collected information of neighboring estimates is considered as new measurements. Therefore, the local and neighboring estimated inputs are assimilated by solving the following set of equations,

$$\begin{bmatrix} \hat{x}_{i,k-1} \\ \hat{u}_{i,k-1} \\ \hat{u}_{j,k-1} \end{bmatrix} = \begin{bmatrix} I & 0 \\ 0 & I \\ 0 & T_{ij} \end{bmatrix} \begin{bmatrix} x^f_{i,k-1} \\ u^f_{i,k-1} \end{bmatrix}, \tag{19}$$

where $\hat{u}_{j,k-1}$ and $P_{uj,k-1}$ are the estimates and covariance of the shared inputs received from $j^{th}$ area, $T_{ij}$ is the transformation matrix so that $u_{j,k-1} = T_{ij} u^f_{i,k-1}$. The covariance matrix of $[\hat{x}_{i,k-1} \ \hat{u}_{i,k-1} \ \hat{u}_{j,k-1}]^T$ is $R_{ij}$, where $R_{ij} = \begin{bmatrix} P_{i,k-1} & 0 \\ 0 & P_{uj,k-1} \end{bmatrix}$. $\begin{bmatrix} x^f_{i,k-1} \\ u^f_{i,k-1} \end{bmatrix}$ represents the assimilated state and input vectors. Given $\mathcal{S} = \begin{bmatrix} I & 0 \\ 0 & I \\ 0 & T_{ij} \end{bmatrix}$, the WLS solution of (19) can be expressed as

$$\begin{bmatrix} \hat{x}^f_{i,k-1} \\ \hat{u}^f_{i,k-1} \end{bmatrix} = (\mathcal{S}^T R_{ij}^{-1} \mathcal{S})^{-1} \mathcal{S}^T R_{ij}^{-1} \begin{bmatrix} \hat{x}_{i,k-1} \\ \hat{u}_{i,k-1} \\ \hat{u}_{j,k-1} \end{bmatrix}. \tag{20}$$

The assimilated covariance matrix become

$$P^f_{i,k-1} = (\mathcal{S}^T R_{ij}^{-1} \mathcal{S})^{-1}. \tag{21}$$

Finally, the state estimation can be continued as the same as (13)-(16) using the assimilated data of $\begin{bmatrix} \hat{x}^f_{i,k-1} \\ \hat{u}^f_{i,k-1} \end{bmatrix}$ and $P^f_{i,k-1}$.

The procedure of the distributed dynamic state-input estimation is summarized as follows.

**Algorithm 2:** Distributed Dynamic State-Input Estimation

**Data Requires**: $z_{xi,k-1}$, $z_{ui,k-1}$, $z_{xi,k}$

1. Solve (12) of local regions for $\hat{x}_{i,k-1}$, $\hat{u}_{i,k-1}$ and $P_{i,k-1}$.

2. Communicate with neighbors for $\hat{u}_{j,k-1}$ and $P_{uj,k-1}$

3. Check Mahalanobis distance (33) for bad data detection

4. Assimilate input estimates by solving (19) with (20)-(21)

5. Update $\hat{x}_{i,k}$ and $P_{i,x,k}$ using (15) and (16)

6. Repeat step 1 at time step $k$

**Remarks**: In traditional WLS and standard Kalam filters, unknown variables or inputs are included in the partitioned state-space model when directly slitting the full one. Owing to the distinct characteristics of the proposed unknown-input dynamic filtering algorithm, it is more advantageous than the traditional ones as unknown inputs are inherently estimated. Although estimation of unknown input causes the proposed method to be more complex in large-scale power systems, the distributed implementation naturally resolves the complexity issue. The local areas can be partitioned internally to perform parallel computing. Thereafter, the common inputs are assimilated to achieve the final estimates.

*C. Bad Data Detection*

Measurements from μPMUs can be highly erroneous due to loss of Global Positioning System (GPS) signals or unexpected faults. This leads to bad data or outliers, which contaminate the estimates. In addition, microgrids and active distribution systems are more and more dependent on information and communication technology (ICT) infrastructure, which is vulnerable to cyber-attacks.

Conventionally, the bad data detection in static SE relies on the residual vector $r_k$, which is the difference between actual and estimated measurements as:

$$r_k = z_k - \hat{z}_k = z_k - H\hat{x}_k, \tag{22}$$

where $z_k$ is the measurement vector; $H$ is the measurement matrix; and $\hat{x}_k$ are the state estimates. However, this approach requires redundant measurements to be able to identify polluted data [27]. Moreover, it suffers from the false data injection attack (FDIA) [28]. When an adversary knows the information of the power networks, or more specific, the measurement matrix $H$, they can bias the estimates by injecting a vector $a_k$ to measurement data as:

$$z_{F,k} = z_k + a_k. \tag{23}$$

The injected vector $a_k$ is built as $a_k = Hx_b$ (24), where $x_b$ is the biased vector. The corrupted estimates $\hat{x}_{a,k}$ obtained by WLS with the weighted matrix $W$ are

$$\hat{x}_{a,k} = (H^TWH)^{-1}H^TW(z_k + Hx_b) = \hat{x}_k + x_b. \tag{25}$$

The residual vector of (22) becomes

$$r_{F,k} = z_k + a_k - H(\hat{x}_k + x_b) = z_k - H\hat{x}_k. \tag{26}$$

Since (22) and (25) are identical, it is impossible to detect such FDIA using the residual vector in static SE.

In Kalman-based DSE, the bad data detection leverages the innovation vector, which is defined as the difference between actual and predicted measurements. In this paper, it is expressed as

$$y_{k-1} = \begin{bmatrix} z_{x,k-1} \\ z_{u,k-1} \\ z_{x,k} \end{bmatrix} - \mathcal{O}\begin{bmatrix} \hat{x}_{k-1} \\ \hat{u}_{k-1} \end{bmatrix}. \tag{27}$$

Equation (27) reflects the consistent between the current measurements and the previous measurements and estimates. According to (12), the innovation vector can be expressed as:

$$y_{k-1} = \mathcal{O} \begin{bmatrix} \tilde{x}_{k-1} \\ \tilde{u}_{k-1} \end{bmatrix} + \begin{bmatrix} v_{x,k-1} \\ v_{u,k-1} \\ Cw_{k-1} + v_{x,k} \end{bmatrix}, \tag{28}$$

where $\tilde{x}_{k-1} = x_{k-1} - \hat{x}_{k-1}$ and $\tilde{u}_{k-1} = u_{k-1} - \hat{u}_{k-1}$. Therefore, if the states and inputs are well-estimated in the previous step and the noise assumption is correct, the innovation vector should have zero means, and the covariance matrix $S_{k-1}$:

$$S_{y,k-1} = \mathcal{O} P_{k-1} \mathcal{O}^T + R_{k-1}. \tag{29}$$

Similarly, the residuals of the input assimilation are expressed as

$$y_{asm,k-1} = \begin{bmatrix} \hat{x}_{i,k-1} \\ \hat{u}_{i,k-1} \\ \hat{u}_{j,k-1} \end{bmatrix} - \begin{bmatrix} I & 0 \\ 0 & I \\ 0 & T_{ij} \end{bmatrix} \begin{bmatrix} \hat{x}^f_{i,k-1} \\ \hat{u}^f_{i,k-1} \end{bmatrix}, \tag{30}$$

and their covariance matrix

$$S_{asm,k-1} = S P^f_{i,k-1} S^T + R_{ij}. \tag{31}$$

**Remarks**: The residual vector in (22) can be expressed as

$$r_k = z_k - M z_k = (I - M) z_k, \tag{32}$$

where $M$ is the projection matrix, $M = H(H^T W H)^{-1} H^T W$ in WLS method. According to (32), adversaries can control the residual vector by manipulating the measurements with knowledge of the measurement matrix $H$. The residual vector reflects only the spatial correlation of measurements at a single time-step. As can be seen in (27), the state and input measurements at two consecutive time-step are employed. Rationally, the innovation vector represents both spatial and temporal correlations by considering the previous measurements. Therefore, leveraging the innovation vector in Kalman-based filtering has higher attack detectability than the residual vector's in static SE [29].

To check the statistics of the innovation vector, it can be processed and put in a statistical test to verify its means and covariances. In this paper, we evaluate the weighted second norm, which is also called the Mahalanobis distance, of the innovation vector. This metric measures the distance between the actual measurement vector $z_k$ and the predicted measurement vector $\hat{z}_k$ as:

$$d_{M,k} = \sqrt{(z_k - \hat{z}_k)^T S^{-1} (z_k - \hat{z}_k)} = \sqrt{y_k^T S^{-1} y_k}, \tag{33}$$

where, $y_k$ is the innovation vector, and $S$ indicates the covariance matrix.

Mahalanobis distance can quantify the dissimilarity between two random vectors of the same distribution with the covariance matrix $S$ [30]. If $S$ is the identity matrix, it becomes the Euclidean distance [31]. Hence, it is more general than the largest normalized residual indicator (LNR) [27], normalized innovation ratio (NIR) [32], and Euclidean distance [31]. For that reason, it is employed as a baseline test in this paper.

*D. Observability Consideration*

The set of equations (12) are overdetermined if and only if

$$rank\,(\mathcal{O}) \geq n + m. \tag{34}$$

Then, (12) can be solved [3]. Therefore, $u_{k-1}$ and $x_{k-1}$ of the time step $(k-1)$ can be estimated using the state measurement of time step $k$. The system becomes observable with one time-step delay.

The total number of measurement devices deploying in the power networks should satisfy the condition (34). Notably, reducing measurements on state variables affects the rank of matrix $\mathcal{O}$ twice compared to input variables. Therefore, we should prioritize measurements of the branch currents over measurements of bus voltages.

Due to the high number of nodes in distribution systems, it is impractical to implement sufficient devices making the entire power networks observable. To improve observability, many studies focused on leveraging the load profiles and smart meter data for *pseudo*-measurements. However, *pseudo*-measurements with low accuracy and high variances may create ill-conditioned matrices during the computing progress [7]. Besides, the slow-updating timescale of RTU- and AMI-based measurements is incompatible with a dynamic estimation algorithm running at a much faster timescale.

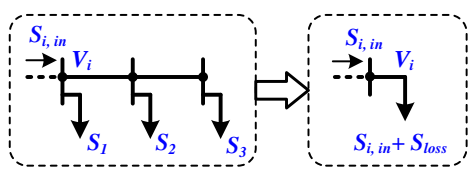

(a) Single-terminal load representation.

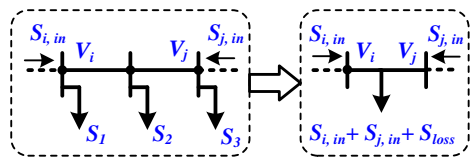

(b) Multiple-terminal load representation.

Fig. 4. Load representation of unobservable areas.

To this end, in this paper, we consider unobservable areas as single- or multiple-terminal loads, as illustrated in Fig. 4. For the state estimation of these areas, the proposed PMU-based dynamic estimation provides the phasor values of $V_i$, $V_j$ and the apparent powers $S_{i,in}$, $S_{j,in}$ at the terminals of these loads. Provided that injection powers, for instance, $S_1$, $S_2$, $S_3$, can be obtained from RTU or AMI-based measurements, the measurement equations of these areas can be expressed as:

$$z_s = h(x_s) + v_s, \tag{35}$$

where $z_s$ is the measurement vector which contains voltage phasors at terminals of loads and the power injections; $x_s$ is the state vector of bus voltages inside; $h(\cdot)$ is the measurement equations; $v_s$ is the measurement errors. Therefore, the inside states can be estimates using traditional SSE technique [33].

The complete state estimation procedure is summarized as

**Algorithm 3:** Complete State Estimation Procedure

**Every 10 ms:** update $z_x$, $z_u$ from PMUs

- Run *dynamic state estimation* algorithm 1 or 2
- Calculate $V_i$, $V_j$,…and $S_{i,in}$, $S_{j,in}$…

**Every 15 minutes:** Update $S_1$, $S_2$, $S_3$… from RTU or AMI-based measurements

- Update $V_i$, $V_j$,…and $S_{i,in}$, $S_{j,in}$…
- Solve (35) for *static estimation*

## IV. CASES STUDY

For the evaluation of the performance of the proposed scheme, a 13-bus, 5-DGUs microgrid, as shown in Fig. 5, representing the Potsdam microgrid in New York State [34], is considered to get simulations results under transient response. The DGUs are inverter-based with the average model, and they are regulated by local primary droop controllers and a secondary control level to attain the nominal voltage and frequency, as shown in Fig. 6. There is the standard primary droop control with the adjustment factors $\Delta V$ and $\Delta f$. These adjustment factors are generated from the secondary control loop, where the frequency of DGU 1 ($f_1$) and the average voltage ($V_{avg}$) of all DGU are regulated following the nominal frequency ($f_{nom}$) and voltage ($V_{nom}$), respectively. The entire Postdam microgrid and all control loops are simulated in real-time on Opal-RT 5700 via MATLAB/Simulink and RT-LAB software. Some load changes are applied to investigate the dynamic response of the system. Time-domain simulation results are sampled at 100 Hz from 0.75 s to 1.75 s. The load changes occur every 0.1 s. The voltage and current data are obtained using sensors deploying in the Potsdam microgrid, as denoted in Fig. 5. The measurements are added white Gaussian noises of $R_u = 5 * 10^{-4}$ pu and $R_i = 5 * 10^{-4}$ pu variances, respectively. These values are chosen based on the measurement precision ($\pm 0.05\%\ TVE$) of PMU. The process noise reflect the uncertainties in the state-transition equation of the state-space model (9). It is the white noise with Gaussian distribution, zero means, and the covariance matrices of $Q = 10^{-4}$ pu, which is reasonably chosen as five-time smaller than the measurement noises.

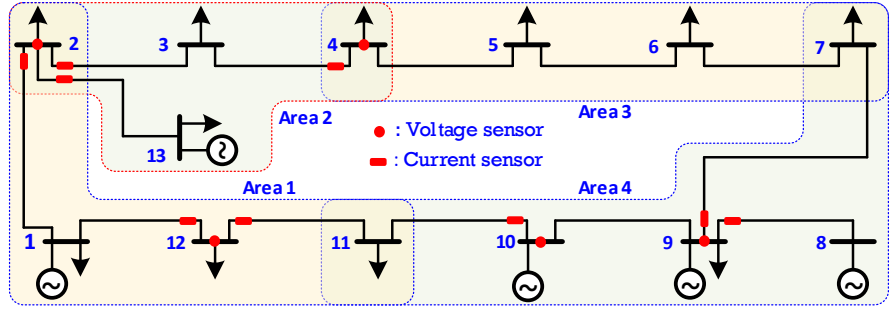

Fig. 5. Schematic of the Potsdam microgrids 13-bus network with four areas.

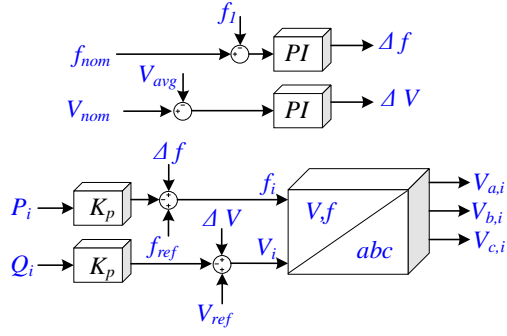

Fig. 6. The block diagram of the hierarchy controls in distributed generation units.

The simulation results compare the proposed state and input estimation (SIE) method with the tracking state estimation (TSE) [5] and traditional WLS-based static estimation (WLS). The true values which are obtained from the simulation model are also added for comparison. To assess the estimation accuracy, the means squared error (MSE) defined as

$$MSE_k = \frac{1}{m}(\hat{u}_k - u_k)^T(\hat{u}_k - u_k), \tag{36}$$

where $k$ is the time index, $\hat{u}_k$ is the estimated values of bus voltages, and $u_k$ is the true value of bus voltages.

There are four areas of the microgrids, wherein each area, the state, and input estimation is performed locally using PMU measurements, except for area 3, which is unobservable and considered as a two-terminal load. The local estimates within area 2 are compared in Section IV-*A*. Thereafter, neighbor estimators communicate with each other to assimilate information of shared inputs and the unobservable area. The assimilated estimates between area 1 and area 4 are compared with the local estimates in Section IV-*B*. Finally, in Section IV-*C*, attack detection is performed, where an attack vector is added to the measurement data. This attack vector is constructed by using equation (24) provided that the matrix $H$ of voltage measurement is known by the adversary.

*A. Local Estimation*

Figs. 7 shows the local estimates of bus voltages 3 in area 2. Here, the upper figure shows the real component $v_{3d}$, while the lower shows the imaginary component $v_{3q}$ of the 3$^{\text{rd}}$ bus voltage. The red lines are the true values of the voltages. The blue ones are the estimates of WLS method. The lines in crimson red are the estimates of the proposed SIE method. The green lines are the estimates of the TSE method. As shown in these figures, these three methods are successful in tracking true values. However, the proposed SIE and WLS methods have better performance since the TSE method converges slower after a load change event. In detail, the TSE method leverages the tracking model of voltages as $v_{i,k+1} = v_{i,k} + w_{v,k}$ (37), whereas the proposed SIE and traditional WLS do not employ this assumption. The tracking model of (37) makes TSE method more reluctant under changes. A forecasting-aided model of $v_{i,k+1} = F_k v_{i,k} + g_k + w_{v,k}$ (38), where $F_k$ is the transition factor and $g_k$ is the trend factor, can be applied [4]. However, under such sudden load changes, the transition and trend factors are unpredictable, then the model (38) still faces the same problem with the model (37).

The MSE of these methods is compared in Fig. 8. It is evident that WLS and SIE have the same MSE, while TSE yields higher MSE under load changes. Notably, although the estimation accuracy of TSE method is higher than that of the WLS and proposed SIE method, the MSE of TSE method is still small and acceptable in general applications. Similarly, Fig. 9 shows the weighted second norms of the residual vector of the three methods. During normal operation, these distances have small values using SIE and WLS methods. The TSE yields higher values, especially under the load changes, since it does not expect a sudden change.

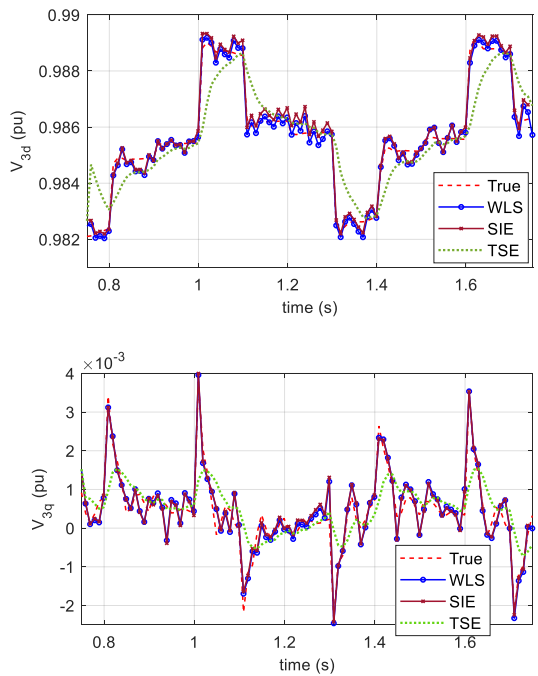

Fig. 7. Estimates of $v_{3d}$ and $v_{3q}$ in areas 2.

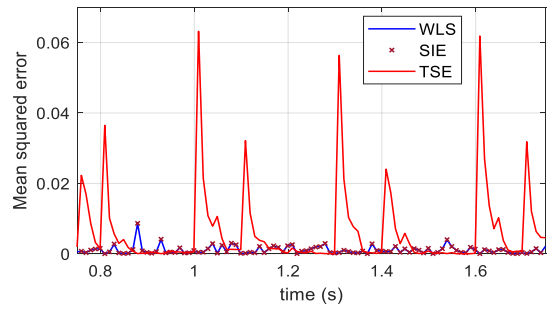

Fig. 8.  Comparison of mean squared errors.

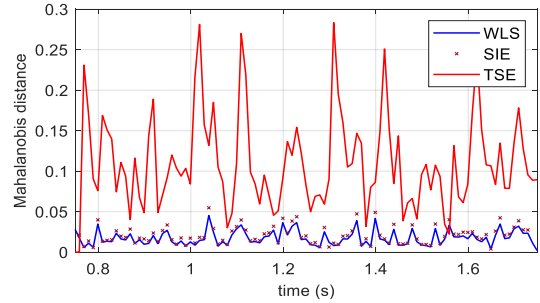

Fig. 9.  Comparison of the weighted second norms of residual vectors.

## B. Assimilated Input Estimates

In this case, the estimated value and covariance of bus voltage 11 are achieved locally at their areas using the proposed SIE method, thereafter these values are transferred between area 1 and area 4 for assimilation. Fig. 10 compares the true value (in red line), the local estimate of area 1 (in green dot) and area 4 (in red star), and the fused estimates (in orange line) of the real component $v_{11d}$ in the upper figure and the imaginary component $v_{11q}$ in the lower figure of the 11$^{th}$ bus voltage. As shown in the figure, the fused and local estimates successfully track the true value, and the fused one is closer to the true value than the local estimates.

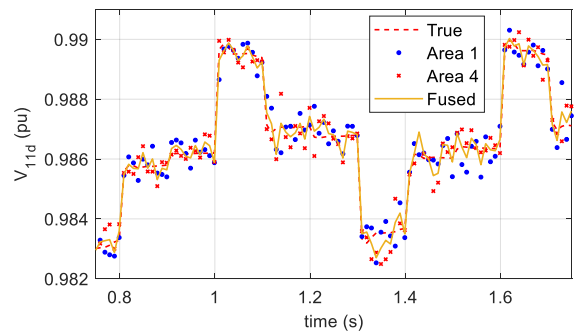

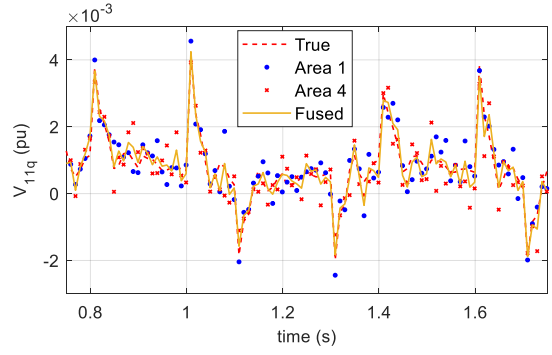

Fig. 10 Estimates of $v_{11d}$ and $v_{11q}$ in areas 1 and 4 with their true values.

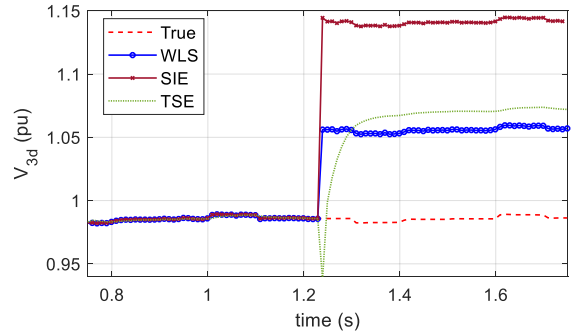

Fig. 11. Estimates of $v_{3d}$ in false data injection attack scenario.

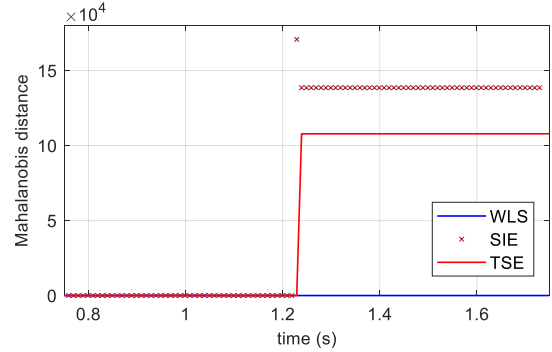

Fig. 12. Comparison of the weighted second norms of residual vectors in false data injection attack scenario.

*C. Attack Detection*

To verify the attack detectability of the proposed method, an attack vector that increases the bus voltages $v_{2,dq}$, $v_{3,dq}$, $v_{4,dq}$, and $v_{13,dq}$ by $x_b = [15 + 0.2i, \ 7.5 + 0.2i, \ 12.5 + 0.2i, \ 8.5 + 0.2i]$ ($\times 10^2 \ V$), respectively, is added to measurement values. This attack is to mislead the control center about the voltage levels. The attack vector is generated using equation (24): $a = Hx_b$, where $H$ is the measurement matrix that maps the measurement values to the targeted state variables. In this attack scenario, the target state variables are the voltages at buses 2, 3, 4, and 13; then the measurements would be the branch currents $I_{2-3}$, $I_{3-4}$, and $I_{13-2}$. The measurement matrix $H$ is expressed as below.

$$\begin{bmatrix} I_{2-3} \\ I_{3-4} \\ I_{13-2} \end{bmatrix} = H \begin{bmatrix} V_2 \\ V_3 \\ V_4 \\ V_{13} \end{bmatrix}, where\ H = \begin{bmatrix} \frac{1}{Z_{23}} & -\frac{1}{Z_{23}} & 0 & 0 \\ 0 & \frac{1}{Z_{34}} & -\frac{1}{Z_{34}} & 0 \\ -\frac{1}{Z_{132}} & 0 & 0 & \frac{1}{Z_{132}} \end{bmatrix}, \quad (37)$$

where $Z_{ij}$ are the line impedances. The measurement values of the branch currents $i_{23,dq}$, $i_{34,dq}$, and $i_{13-2,dq}$ are recorded and then are added with the attack vector $a = Hx_b = [0.2421 - 0.2994i, -5.3585 + 6.6276i, -0.2098 + 0.2595i]\ (\times 10^4\ A)$ for a false measurement data vector before it was sent to the control center for necessary control actions.

Fig. 11 shows the true value and the estimates of the real component of the 3rd bus voltage $v_{3d}$ using the proposed SIE method, the WLS. The attack scenario is triggered at 1.25 s, the estimate values are immediately contaminated, going far from the true value.

Fig. 12 shows the weighted second norms of residual vectors of three methods under the attack. It is evident that the WLS method is unable to recognize the attack since its distance does not change under attack. The SIE and TSE methods show a big leap in the weighted second norms, which indicates bad data, or an attack happens. Therefore, this FDIA can be detected with SIE and TSE algorithms.

## V. CONCLUSION

This paper proposes a dynamic estimation scheme with unknown inputs for power networks in microgrids and active distribution networks supporting by µPMU measurements. To the best of author's knowledge, this is the first work on simultaneous input and state dynamic estimation applied in power systems. The differential equations of branch currents are employed to build a decoupled dynamic model of power networks, where the branch current is the state vector, and the bus voltages are the unknown input vector. To simultaneously estimate both state and input vectors, modified linear Kalman-based filtering algorithms with unknown input variables are proposed. The distributed implementation of the proposed method is also presented. The case studies compared the proposed method with the traditional weighted least squares and tracking state estimation method under normal operations and attack cases. The simulation results have demonstrated the estimation performance of the proposed method. The mean square errors are smaller than 0.01 and the Mahalanobis distances are smaller than 0.05 under normal operation. The proposed method is able to detect false data injection attacks by evaluating the Mahalanobis distance. This parameter metric is larger than $10^5$ under attacks in the case studies.

## APPENDIX: MICROGRID MODEL PARAMETERS

a.  *Cable information:* The line parameters are taken from the real Potsdam microgrid [34] and listed here as follows:

$l_{1-2} = 3100$, $l_{2-3} = 4150$, $l_{3-4} = 125$, $l_{4-5} = 3350$, $l_{5-6} = 4350$, $l_{6-7} = 5425$, $l_{7-9} = 7025$, $l_{8-9} = 8100$, $l_{9-10} = 8200$, $l_{10-11} = 375$, $l_{11-12} = 400$, $l_{1-12} = 1950$, $l_{2-13} = 4150$.

Note that the line lengths (in feet). The cable type is 500 MCM with positive sequence resistance of 0.1558 $\Omega/mi$, the reactance of 0.1927 $\Omega/mi$, and susceptance of 253.54 $\mu S/mi$.

b. *Loading information:* The loads are simulated using a three-phase dynamic load block in MATLAB/Simulink with nominal values of active power (MW) and reactive power (MVA) at a nominal voltage (13.2 kV) listed as follows:

$P_1 = 4866, Q_1 = 3015, P_2 = 48, Q_2 = 30, P_3 = 144, Q_3 = 89, P_4 = 54, Q_4 = 33, P_5 = 560, Q_5 = 347, P_6 = 122, Q_6 = 76, P_7 = 142, Q_7 = 88, P_9 = 4166, Q_9 = 2582, P_{11} = 48, Q_{11} = 30, P_{12} = 48, Q_{12} = 30, P_{13} = 83, Q_{13} = 51.$

c. *Distributed generation unit:* The distributed generation units are the inverter-based average models where the controlled voltage source blocks are employed. The internal resistances ($\Omega$) & inductances ($mH$) are added to represent the internal RL filter of inverters, the connecting lines to the point of common coupling, and step-up transformers, listed as follows:

$R_{t1} = 0.3, L_{t1} = 7.8, R_{t8} = 0.4, L_{t8} = 10, R_{t9} = 0.2, L_{t9} = 4.4, R_{t10} = 1, L_{t10} = 26, R_{t13} = 1.2, L_{t13} = 29.$

These impedances were designed properly to ensure microgrid stability.

## ACKNOWLEDGEMENT

This material is based upon research supported by, or in part by, the U.S. Office of Naval Research under award number N00014-16-1-2956.